\documentclass[a4paper]{article}

\usepackage{INTERSPEECH2019}
\usepackage{booktabs}

\title{CSTNet: Contrastive Speech Translation Network for Self-Supervised Speech Representation Learning}
\name{\small Sameer Khurana$^1$, Antoine Laurent$^2$, James Glass$^1$}
\address{
  $^1$MIT Computer Science and Artificial Intelligence Laboratory, Cambridge, MA, USA\\
  $^2$LIUM - Le Mans University, France}
\email{\{skhurana,glass\}@mit.edu, antoine.laurent@univ-lemans.fr}

\begin{document}

\maketitle
\begin{abstract}
More than half of the 7,000 languages in the world are in imminent danger of going extinct. Traditional methods of documenting language proceed by collecting audio data followed by manual annotation by trained linguists at different levels of granularity. This time consuming and painstaking process could benefit from machine learning. Many endangered languages do not have any orthographic form but usually have speakers that are bi-lingual and trained in a high resource language. It is relatively easy to obtain textual translations corresponding to speech. In this work, we provide a multimodal machine learning framework for speech representation learning by exploiting the correlations between the two modalities namely speech and its corresponding text translation. Here, we construct a convolutional neural network audio encoder capable of extracting linguistic representations from speech. The audio encoder is trained to perform a speech-translation retrieval task in a contrastive learning framework. By evaluating the learned representations on a phone recognition task, we demonstrate that linguistic representations emerge in the audio encoder's internal representations as a by-product of learning to perform the retrieval task.
\end{abstract}

\section{Introduction}

UNESCO’s ``Atlas Of The World’s Languages In Danger” marks 43\% of the languages in the world  as endangered. It has been  argued that at the current rate of extinction, more than 90\% of the world's languages will disappear in the next hundred years. Loss of a language leads to loss of cultural identity, the loss of linguistic diversity and in general, loss of knowledge. There are many reasons for a language to become endangered as mentioned in \cite{anastasopoulos2019computational}. The steps taken for documenting endangered languages is quite painstaking. It includes codifying the rules governing the language by trained linguists at different levels such as phonetics, phonology, morphology, syntax and so on. In order to facilitate language documentation the data collected by field linguists consists of speech data, its orthographic transcription, if available, and spoken/written translation in a high resource language. For many languages no orthographic form exists. Nevertheless, it is relatively easy to provide written or spoken translations for audio sources, as speakers of a minority language are often bilingual and literate in a high-resource language \cite{adda2016breaking}. Hence, oftentimes the only textual information that is available for endangered languages is in the form of translations. This paired speech-translation data source can be exploited by machine learning algorithms to build systems of linguistic structure discovery for speech in endangered language, as shown by the excellent work presented in \cite{anastasopoulos2019computational}. In this work, we present the Contrastive Speech Translation Network (CSTNet), a deep learning based multi-modal framework, that learns low-level linguistic representations for speech by exploiting the paired speech-translation data source.

Our CSTNet is inspired by the multimodal \textbf{D}eep \textbf{A}udio-\textbf{V}isual \textbf{E}mbedding \textbf{N}etwork (DAVENet) and subsequent ResDAVENet of Harwath et al.~\cite{harwath2016unsupervised,harwath2019jointly}, along with their more recent research on sub-word unit learning within the DAVENet and ResDAVENet models~\cite{harwath2019towards,harwath2020learning}. They propose neural models of Audio-Visual grounding, where they construct neural network models that learn by associating spoken audio captions with their corresponding image. Their framework consists of an audio encoder and an image encoder, both parametrized using Deep Neural Networks. Association between the spoken audio captions and their corresponding image are learned by using a constrastive learning framework which is a triplet loss between the embeddings outputted by the audio and the image encoders. They show that by performing the speech-image retrieval task, linguistic representations emerge in the internal representation of the audio encoder. In this work, we reach similar conclusion by performing the task of speech-translation retrieval using the aforementioned contrastive learning framework. We give details about our model in Section~\ref{sec:cstnet}.

As a proof of concept, we train the CSTNet on speech-translation pairs where the speech side is always English and the text translation side is either French, German, Spanish, Portugese or Italian. We obtain this paired dataset from the Multilingual Speech Translation Corpus (MuST-C) (Section~\ref{sec:dataset}). In this work, we make the following \textbf{contributions}:
\begin{itemize}
\item We present a self-supervised learning \cite{de1994learning} framework for linguistic representation learning from speech, without any manual labels, that learns by performing the task of speech-translation retrieval. This framework has the potential to be used in documenting endangered language where such speech-translation paired data exists. Besides language documentation, this is also a novel self-supervised learning framework for speech representation learning.
\item We analyze the representations learned by the CSTNet's audio encoder on minimal-pair ABX task \cite{schatz2013evaluating} proposed as part of the Zero Resource Speech Challenge \cite{dunbar2019zero}. We show that our model outperforms the best system, based on Wavenet-VQ \cite{chorowski2019unsupervised}, by 8 percentage points (pp) and is comparable to the recently proposed ResDAVENet \cite{harwath2020learning}. In addition, we show that the representations learned by the CSTNet encodes phonetic information as evidenced by the good performance on the downstream phone classification task on the Wall Street Journal dataset.
\end{itemize}

\section{Related work}
Unsupervised learning methods can be categorized into self-supervised learning methods and generative models. Recently, several self-supervised learning methods have been proposed that learn from only speech data only. Methods like Problem Agnostic Speech Encoder (PASE) \cite{pascual2019learning}, MockingJay \cite{liu2020mockingjay}, Wav2Vec \cite{schneider2019wav2vec} and Autoregressive Predictive coding (APC) \cite{chung2019unsupervised} fall into this category. Wav2Vec is a Convolutional Neural Network (CNN) based contrastive predictive learning framework. A CNN audio encoder provides low-frequency representations from raw waveforms and the model learns by maximizing mutual information between the past and future feature representations output by the encoder. PASE is multi-task framework that uses a SincNet \cite{ravanelli2018speaker} encoder to embed a raw waveform into a continuous representation. The encoder is trained by performing multiple prediction and mutual information maximization tasks conditioned on the audio embedding output by the SincNet encoder. APC and MockingJay borrow the self-supervised learning methods proposed in the field of Natural Language Processing (NLP). APC constructs a spectrogram language model using a Recurrent Neural Network (RNN). The model is trained on the future frame prediction task, conditioned on the past information, by minimizing the L1 loss between the predicted and the ground truth acoustic frame. MockingJay performs the task of masked self-prediction of the raw Mel-spetrogram. They use a Transformer encoder inspired by the BERT \cite{devlin2018bert} architecture in NLP. So far, the self-supervised learning approaches we discussed use only speech data. Another important class of self-supervised methods are the Deep Audio-Visual Embedding Networks (DAVENet) \cite{harwath2016unsupervised, harwath2019towards, harwath2020learning}. They train a CNN audio encoder that learns to associate spoken captions with its corresponding image. They show that by training the network to perform the speech image retrieval task, the internal representations of audio network learn linguistic representations as a by-product. As far as we know, \textbf{our work is the first one} in which the two modalities are speech data and it's textual translation in different languages.

Generative models have seen renewed interest over the past years due to the introduction of Variational Autoencoders (VAEs) \cite{kingma2013auto}.  VAEs have been used for disentangled representation of speech \cite{hsu2017unsupervised, khurana2019factorial, li2018disentangled}. Besides VAE, Autoregressive models, a class of explicit density generative models, have been used to construct speech density estimators. Neural Autoregressive Density Estimatior (NADE) \cite{uria2016neural} is a prominent earlier work followed by more recent WaveNet \cite{oord2016wavenet}, SampleRNN \cite{mehri2016samplernn} and MelNet \cite{vasquez2019melnet}. An interesting avenue of future research is to probe the internal representations of these models for linguistic information. We note that WaveGlow, a flow based generative model, has been recently proposed as an alternative to autoregressive models for speech \cite{prenger2019waveglow}. Generative adversarial networks (GANs), an implicit density generative model, have also been used to model speech \cite{donahue2018adversarial}. Autoregressive generative models and WaveGlow are generally used as Vocoders for speech synthesis in Text-to-Speech synthesis systems. It is not clear how to use these systems for representation learning.

\begin{figure*}[h]
    \centering
    \includegraphics[width=\linewidth]{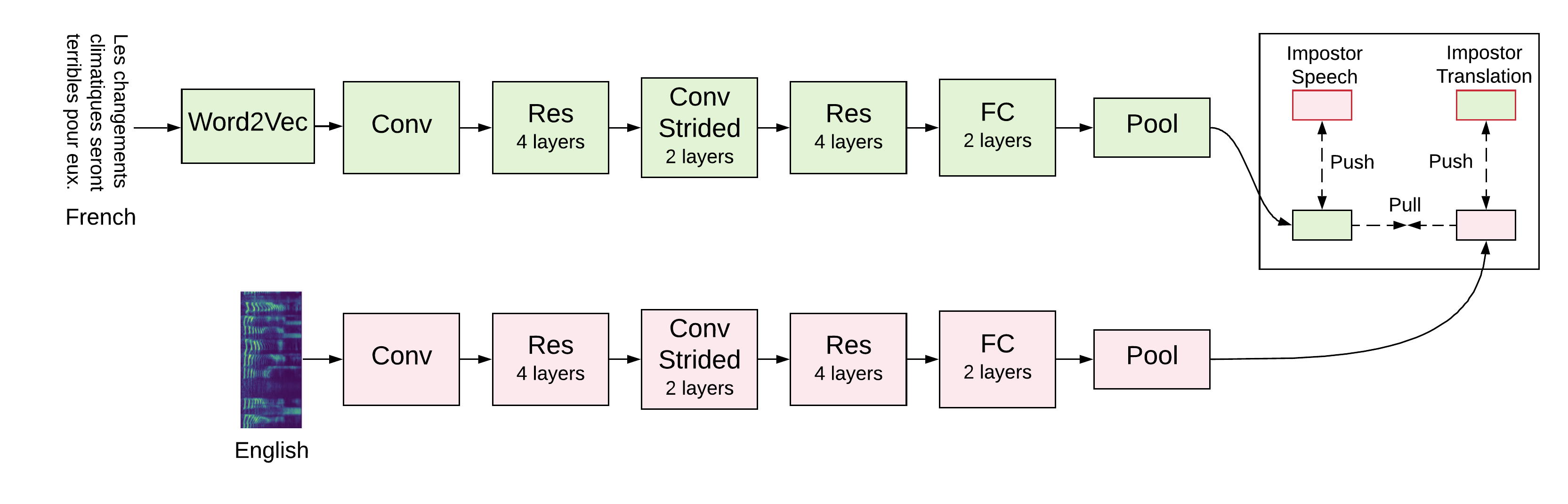}
    \caption{Diagram of the Contrastive Speech Translation Network.}
    \label{fig:model}
\end{figure*}


\section{Dataset}
\label{sec:dataset}
We use the freely downloadable MuST-C \cite{di-gangi-etal-2019-must} corpus, a multilingual speech translation corpus, to train our model.
For each of the 8 languages targeted by MUST-C, the corpus contains at least 385 hours of audio recordings from English TED talks, which are automatically aligned at the sentence level with their manual transcriptions and translations.
Statistics of the corpus that we used (5 languages out of 8) are given Table \ref{tab:statmust} extracted from \cite{di-gangi-etal-2019-must}.

\begin{table}[h]
    \caption{Statistics of the MUST-C corpus}
    \centering
    \begin{tabular}{c|c|c|c}
       \textbf{Tgt} & \textbf{\#Talk} & \textbf{\#Sent} & \textbf{Hours}\\
       \hline
        \textbf{De} & 2093 & 234k & 408 \\
        \textbf{Es} & 2564 & 270k & 504 \\
        \textbf{Fr} & 2510 & 280k & 492 \\
        \textbf{It} & 2374 & 258k & 465 \\
        \textbf{Pt} & 2050 & 211k & 385 \\
    \end{tabular}
    
    \label{tab:statmust}
\end{table}

The corpus was also divided into development, test and train sets.
The test and development corpus is built with segments from talks that are common to all the languages. Their size is, respectively, 1.4K (from 11 talks) and 2.5K segments (from 27 talks). The remaining data (of variable size depending on the language pairs) are used for training.

\section{Contrastive Speech Translation Network}
\label{sec:cstnet}
\subsection{Neural model}

As illustrated in Figure~\ref{fig:model}, our model consists of two embedding functions that embed audio and text sequences into a fixed dimensional vector. The embedding functions are parametrized as an 11-layered convolutional neural network (CNN) with residual connections, followed by 2 fully-connected layers and a mean pooling layer at the end that gives a fixed dimensional embedding. The first layer of the network is a 1-D convolution that spans the entire spatial dimension of the input signal, while the remaining 10 1-D convolution layers are across the time axis. The 10 layers are divided into 2 residual blocks of 4 layers each, interleaved with two strided convolution layers with stride of 2. We also use Batch-Normalization \cite{ioffe2015batch} to normalize the activations after each hidden layer in the network. Finally, the output of the convolution layers is mean pooled across the time axis, to give a single embedding vector that represents the input feature sequence. Both the audio and the text embedding functions use the same CNN architecture. We use 1024 hidden channels for each layer and hence the size of the output embedding vector is 1024. The CNN architecture is inspired by the audio encoder presented in Chorowski et al. \cite{chorowski2019unsupervised}. 

The input to the audio network is 40 dimensional Mel-FBanks extracted with a Hamming window of size 25 ms and stride 10 ms. The input to the text network is the sequence of word embeddings that make up the sentence. To extract word embeddings we use the pre-trained Word2Vec models for different languages publicly available through the FastText library. This gives 100 dimensional word embeddings. There is clearly opportunity to consider word embeddings extracted from pre-trained Language Models such as BERT \cite{devlin2018bert}, GPT-2 \cite{radford2019language} etc. We leave this line of investigation for future work.

\subsection{Triplet loss training}
Our model was trained using the same loss function as \cite{harwath2019towards, harwath2020learning}.
This loss function is a mix of two triplet loss terms \cite{bredin2017tristounet,weinberger2009distance}, one based on random sampling of negative examples, and the other based on semi-hard negative mining, in order to find more challenging negative samples. Below we give in some detail about how the triplet loss is computed. This formulation is taken from Harwath et. al \cite{harwath2020learning}.

Given two sets of output embeddings, $\mathbb{A}=\{a_1, ..., a_B\}$ and $\mathbb{T}=\{t_1, ..., t_B\}$, in  a batch size of $B$ audio/translation training pairs, the randomly-sampled triplet loss term is computed by randomly selecting impostor examples, $\bar{a}_j$ and $\bar{t}_j $ for the  $j_{th}$ element in the batch and then computing the randomly-sampled triplet loss as follows:
\begin{equation}
    \mathcal{L}_s = \sum_{j=1}^{B}(\textrm{max}(0, t_j^T\bar{a}_j-t_j^Ta_j+1)+\textrm{max}(0,\bar{t}_j^Ta_j-t_j^Ta_j+1))
\end{equation}

For the semi-hard negative triplet loss, we first define the sets of impostors candidates for the $j_{th}$ example as $\hat{\mathbb{A}}_j = \{a \in \mathbb{A} | t_j^Ta < t_j^Ta_j\}$ and $\hat{\mathbb{T}}_j=\{t \in \mathbb{T} | t^Ta_j < t_j^Ta_j\}$. The semi-hard negative loss is then computed as:
\begin{align*}
    \mathcal{L}_h = \sum_{j=1}^B(\textrm{max}(0,\underset{\hat{a} \in \hat{\mathbb{A}}_j}{\textrm{max}}(t_j^T\hat{a})-t_j^Ta_j+1) + \\
    \textrm{max}(0, \underset{\hat{t} \in \hat{\mathbb{T}}_j}{\textrm{max}}(\hat{t}^Ta_j)-i_j^T+1))
\end{align*}

Finally, the overall loss function is computed by combining the two above losses, $\mathcal{L} = \mathcal{L}_s + \mathcal{L}_h$, which was found by \cite{harwath2019towards} to outperform either loss on its own.

The model is trained using the Adam optimizer with a learning rate of 0.001 for 100 epochs. We decay the learning rate by multiplying it with a factor of 0.95 every three epochs. L2 regularization on model parameters with weight 5e-7 is used during training. 

\section{Experiments and results}
\subsection{Evaluation Protocol and Dataset}
We evaluate the internal representations learned by the CSTNet's audio encoder on two tasks: the Minimal-Pair ABX (MP-ABX) task \cite{versteegh2016zero, dunbar2017zero} and phone recognition. MP-ABX provides an unsupervised and non-parametric way of evaluating speech representations. It measures ABX-discriminability between phoneme triples that differ only in their center phoneme (for example for phonemes `bed' and `bad'). For phoneme triples $a$ and $x$ from the same category ($A$) and $b$ from another category ($B$), the ABX-discriminability in the ZeroSpeech challenge is defined as the probability that the Dynamic Time Warping (DTW) divergence between $a$ and $x$ is smaller than that between $b$ and $x$. ABX performance is tested on the ZeroSpeech Challenge 2019 (ZRC19) English test set. For phone recognition, we pass the features through a softmax layer that is trained using Connectionist Temporal Classification (CTC) to predict the output phone sequence. By following this protocol for phone recognition, we ensure that the task performance is solely driven by the learned representations. For phone recognition, the softmax classifier is trained on 80 hours of the Wall Street Journal (WSJ) train dataset and evaluated on the WSJ eval92 dataset. We do not fine tune the pre-trained audio network on downstream tasks. We also present speech translation retrieval performance of CSTNet. This shows how well the model is performing on the actual task that it is trained on.

Features for the downstream tasks are extracted from different layers of the pre-trained CSTNet which is trained on different speech-translation pairs of the MuST-C corpus.

\subsection{Results and Discussion}
In Table~\ref{tab:retrieval}, we present the speech translation retrieval using the recall accuracy from text to speech and speech to text. This gives us an indication of how well CSTNet is doing at the actual task that it is trained on. Rows correspond to retrieval performance for the model trained on different language pairs of the MuST-C corpus.
\begin{table}[h]
    \caption{Experimental results for speech to text and text to speech retrieval task.}
    \label{tab:retrieval}
    \centering
    \resizebox{\linewidth}{!}{
    \begin{tabular}{lccccccc}\toprule
    & \multicolumn{3}{c}{\textbf{Speech $\rightarrow$ Text}}&&\multicolumn{3}{c}{\textbf{Text $\rightarrow$ Speech}}\\
    \cmidrule{2-4}\cmidrule{6-8}
    \textbf{Language pair}&R@10&R@5&R@1&&R@10&R@5&R@1\\
    \hline\\
    en-fr &75.4&67.9&43.5&&72.5&67.1&29.0\\
    en-de &73.9&64.4&38.8&&66.9&61.6&26.6\\
    en-es &79.9&73.3&49.6&&77.2&73.3&37.2\\
    en-it &72.7&62.9&38.7&&67.8&61.7&27.8\\
    en-pt &69.4&59.8&36.3&&64.6&58.2&25.4\\
    \bottomrule
    \end{tabular}
    }
\end{table}

\begin{figure}[h]
    \centering
    \includegraphics[width=\linewidth]{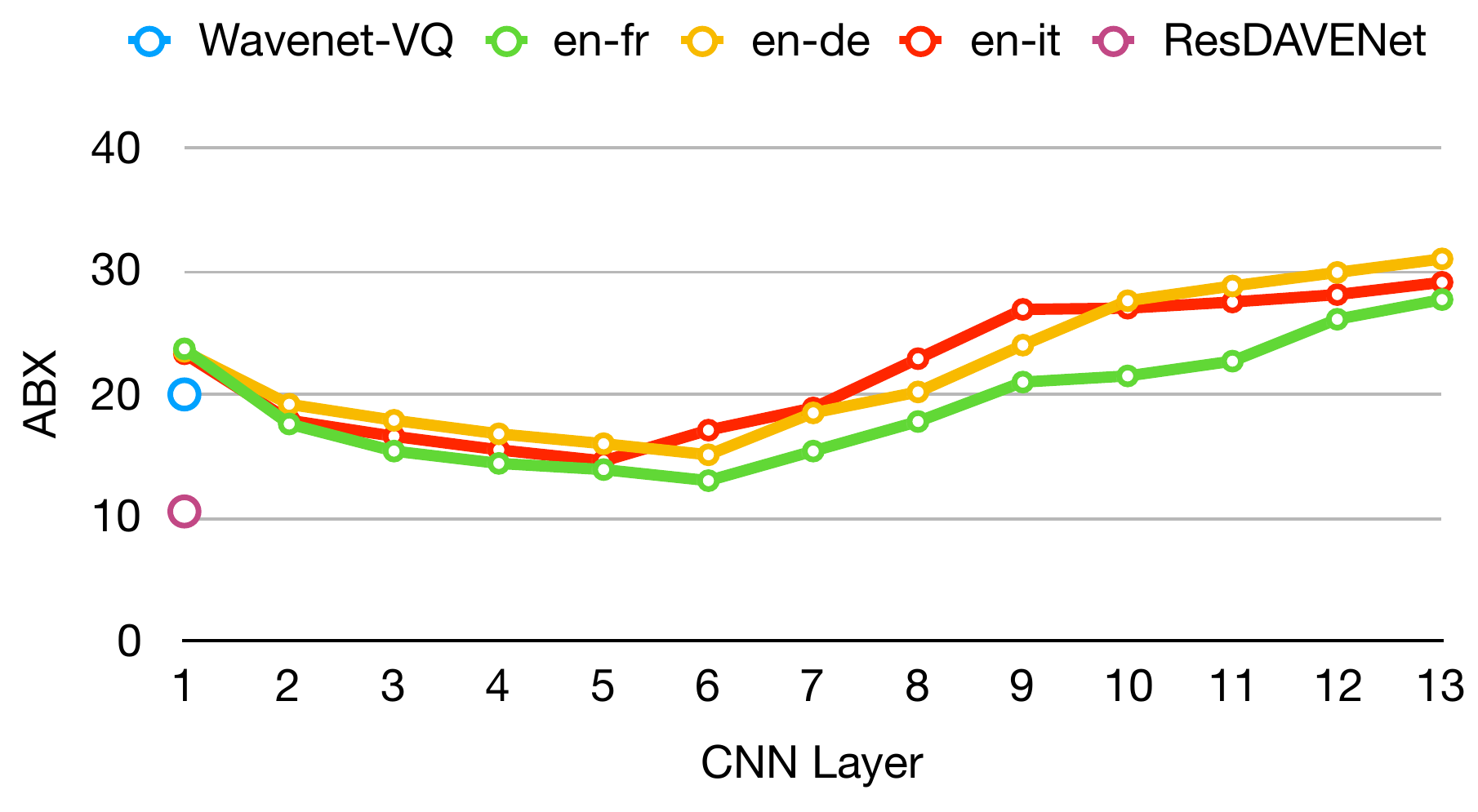}
    \caption{ABX error vs Audio Network's layer number}
    \label{fig:abx}
\end{figure}

In Table~\ref{tab:abx}, we present the best ABX scores (lower is better), on the ZRC19 English test set. We compute the ABX score using all the layers of the pre-trained CSTNet's audio encoder trained on different language pairs. Here, we present the best results that is usually obtained using the representations in the middle of the network (for layer numbers 5, 6). In Figure~\ref{fig:abx}, we show the curve of ABX scores vs audio network layer number for the CSTNet trained on three different language pairs. The network hits a sweet spot in the middle layers, where the receptive field is approximately 100-140 ms. A similar trend is observed for all the languages. We significantly outperform Wavenet-VQ (ZS), the best performing  submission to the ZRC19 challenge, based on Vector Quantization VAE (VQ-VAE) \cite{chorowski2019unsupervised}, that is trained on the ZeroSpeech (ZS) training set. To have a fair comparison, we also compare our model against Wavenet-VQ (MuST-C) that is trained on the English speech from the MuST-C corpus on which CSTNet is also trained. CSTNet still outperforms Wavenet-VQ. Hence, we show that our framework could be an alternative to reconstruction based representation learning methods. As compared to the best reported ResDAVENet model in \cite{harwath2020learning}, the audio visual system, our best model, trained on English-Spanish language pair, lags behind by 1.5 points.

\begin{table}[h]
    \caption{ABX Scores on ZRC19 Challenge English Test Set}
    \label{tab:abx}
    \centering

    \begin{tabular}{l|c}
    \textbf{Method} & \textbf{ABX} \\
    \hline
    ResDAVENet & 10.8 \\
    en-es & 12.3 \\
    en-fr & 13.0 \\
     en-it & 14.6 \\
    en-de & 15.1 \\
    en-pt & 16.7 \\
    Wavenet-VQ (ZS) & 19.9 \\
    Wavenet-VQ (MuST-C) & 20.1 \\
    \end{tabular}
\end{table}

In Table~\ref{tab:1}, we analyze the representations learned by different layers of the CSTNet's audio encoder. Rows in the table correspond to the CSTNet trained on different language pairs and the different columns correspond to the layer number of the pre-trained audio network. As can be seen from the table, as we go higher up in the audio network, the phone recognition performance improves with the best performance achieved using the representations extracted from the last layer. The improvement in PER is not consistent with the increase in layer size but the best performance is always from last layer. This is not surprising as the last layer has the highest receptive field and hence has access to more global information essential for the task of phone recognition. The best performance is obtained using the CSTNet trained on the language pair English-French, but the performance gap between different language pairs is not very significant, with a gap of 2.8 percentage points between the best (en-fr) and the worst (en-de) language pairs.
\begin{table}[h]
    \caption{Phone Recognition Results with features extracted from different layers of CSTNet on WSJ eval92. LX stands for Layer \#X.}
    \label{tab:1}
    \centering
    \resizebox{\linewidth}{!}{
    \begin{tabular}{l|ccccccccc|}
     &  \multicolumn{9}{c|}{\textbf{PER}} \\
    \textbf{Language pair} & L5& L6 & L7 & L8 & L9 & L10 & L11 & L12 & L13 \\
    \hline
    en-fr  &40.9&34.5&33.5&32.1&31.9&32.5&32.5&32.5&29.2 \\
    en-de &52.8&38.6&40.9&41.4&41.1&43.4&41.1&38.3&32.5 \\
    en-es &42.7&34.8&33.5&33.4&33.2&33.7&34.9&35.4&30.3 \\
    en-it &82.8&25.5&34.4&34.6&33.7&35.7&36.0&35.8&30.9 \\
    en-pt &83.3&38.3&37.9&38.1&39.3&38.7&42.1&40.0&31.7 \\
    \end{tabular}
    }
\end{table}

In Table~\ref{tab:comp}, we compare the CSTNet with other self-supervised learning systems on the task of phone recognition. Representations learned by CSTNet significantly outperform all other systems except Wav2Vec, where the best CSTNet system trained on English-French lags by 8 percentage points (pp). Our worst model, English-German, outperforms ResDAVENet by 6pp. For ResDAVENet, we compute the PER using features from every layer and report the best results. This is an encouraging result showing that CSTNet can learn better linguistic information than the ResDAVENet trained on the audio-visual retrieval task. We do not train any of the comparison features extractors on our own, but use the publicly available checkpoints released by the authors of the respective methods.
\begin{table}[h]
    \caption{Phoneme Error Rate using multiple self-supervised learning methods.}
    \label{tab:comp}
    \centering

    \begin{tabular}{l|c}
    \textbf{Method} & \textbf{PER} \\
    \hline
    Wav2Vec \cite{schneider2019wav2vec} & 21.6\\
    en-fr & 29.2 \\
    en-es & 30.3 \\
    en-it & 30.9 \\
    en-pt & 31.7 \\
    en-de & 32.5 \\
    ResDAVENet \cite{harwath2020learning} &  38.5\\
    MockingJay \cite{liu2019mockingjay} & 41.2 \\
    PASE \cite{pascual2019learning} & 45.2\\
    \end{tabular}
\end{table}

We acknowledge that in this work we have not shown the usefulness of our framework as part of any real world linguistic annotation toolkit used for documenting endangered language, nonetheless, we argue that by empirically demonstrating the capability of CSTNet to acquire linguistic information in its internal representations, it could form an integral part of the linguistic structure discovery systems. It could be composed with the non-parametric Bayesian model of Acoustic Unit Discovery of Lee \& Glass \cite{lee2012nonparametric}, Variational AUD system of Ondel et. al \cite{ondel2016variational} and Ebbers et. al \cite{ebbers2017hidden} where the CSTNet audio encoder will play the role of feature extractor.

\section{Conclusions and Future Work}

In this paper, we propose the Contrastive Speech Translation Network (CSTNet), a self-supervised learning framework for learning linguistic representations from speech using the speech translation retrieval task. To the best our knowledge, this is the first work that uses paired speech-translation data for speech representation learning. We show that the speech representations learned by our framework outperformed multiple representation learning systems on the downstream task of phone recognition. In the future, we would apply our proof of concept to a real world language documentation task. Another interesting future direction is to learn not just phonetic information but also sub-word and word like information using this framework. To that end, we would follow the work on learning discrete hierarchical units presented in ResDAVENet-VQ \cite{harwath2020learning}, where the authors use interleaved Vector Quantization layers in the audio network of their audio-visual retrieval system.




\bibliographystyle{IEEEtran}

\bibliography{mybib}

\begin{thebibliography}{10}
\providecommand{\url}[1]{#1}
\csname url@samestyle\endcsname
\providecommand{\newblock}{\relax}
\providecommand{\bibinfo}[2]{#2}
\providecommand{\BIBentrySTDinterwordspacing}{\spaceskip=0pt\relax}
\providecommand{\BIBentryALTinterwordstretchfactor}{4}
\providecommand{\BIBentryALTinterwordspacing}{\spaceskip=\fontdimen2\font plus
\BIBentryALTinterwordstretchfactor\fontdimen3\font minus
  \fontdimen4\font\relax}
\providecommand{\BIBforeignlanguage}[2]{{%
\expandafter\ifx\csname l@#1\endcsname\relax
\typeout{** WARNING: IEEEtran.bst: No hyphenation pattern has been}%
\typeout{** loaded for the language `#1'. Using the pattern for}%
\typeout{** the default language instead.}%
\else
\language=\csname l@#1\endcsname
\fi
#2}}
\providecommand{\BIBdecl}{\relax}
\BIBdecl

\bibitem{anastasopoulos2019computational}
A.~Anastasopoulos, \emph{Computational Tools for Endangered Language
  Documentation}.\hskip 1em plus 0.5em minus 0.4em\relax University of Notre
  Dame, 2019.

\bibitem{adda2016breaking}
G.~Adda, S.~St{\"u}ker, M.~Adda-Decker, O.~Ambouroue, L.~Besacier, D.~Blachon,
  H.~Bonneau-Maynard, P.~Godard, F.~Hamlaoui, D.~Idiatov \emph{et~al.},
  ``Breaking the unwritten language barrier: The bulb project,'' \emph{Procedia
  Computer Science}, vol.~81, pp. 8--14, 2016.

\bibitem{harwath2016unsupervised}
D.~Harwath, A.~Torralba, and J.~Glass, ``Unsupervised learning of spoken
  language with visual context,'' in \emph{Advances in Neural Information
  Processing Systems}, 2016, pp. 1858--1866.

\bibitem{harwath2019jointly}
D.~Harwath, A.~Recasens, D.~Sur{\'\i}s, G.~Chuang, A.~Torralba, and J.~Glass,
  ``Jointly discovering visual objects and spoken words from raw sensory
  input,'' \emph{International Journal of Computer Vision}, pp. 1--22, 2019.

\bibitem{harwath2019towards}
D.~Harwath and J.~Glass, ``Towards visually grounded sub-word speech unit
  discovery,'' in \emph{ICASSP 2019-2019 IEEE International Conference on
  Acoustics, Speech and Signal Processing (ICASSP)}.\hskip 1em plus 0.5em minus
  0.4em\relax IEEE, 2019, pp. 3017--3021.

\bibitem{harwath2020learning}
\BIBentryALTinterwordspacing
D.~Harwath, W.-N. Hsu, and J.~Glass, ``Learning hierarchical discrete
  linguistic units from visually-grounded speech,'' in \emph{International
  Conference on Learning Representations}, 2020. [Online]. Available:
  \url{https://openreview.net/forum?id=B1elCp4KwH}
\BIBentrySTDinterwordspacing

\bibitem{de1994learning}
V.~R. de~Sa, ``Learning classification with unlabeled data,'' in \emph{Advances
  in neural information processing systems}, 1994, pp. 112--119.

\bibitem{schatz2013evaluating}
T.~Schatz, V.~Peddinti, F.~Bach, A.~Jansen, H.~Hermansky, and E.~Dupoux,
  ``Evaluating speech features with the minimal-pair abx task: Analysis of the
  classical mfc/plp pipeline,'' 2013.

\bibitem{dunbar2019zero}
E.~Dunbar, R.~Algayres, J.~Karadayi, M.~Bernard, J.~Benjumea, X.-N. Cao,
  L.~Miskic, C.~Dugrain, L.~Ondel, A.~W. Black \emph{et~al.}, ``The zero
  resource speech challenge 2019: Tts without t,'' \emph{arXiv preprint
  arXiv:1904.11469}, 2019.

\bibitem{chorowski2019unsupervised}
J.~Chorowski, R.~J. Weiss, S.~Bengio, and A.~van~den Oord, ``Unsupervised
  speech representation learning using wavenet autoencoders,'' \emph{IEEE/ACM
  transactions on audio, speech, and language processing}, vol.~27, no.~12, pp.
  2041--2053, 2019.

\bibitem{pascual2019learning}
S.~Pascual, M.~Ravanelli, J.~Serr{\`a}, A.~Bonafonte, and Y.~Bengio, ``Learning
  problem-agnostic speech representations from multiple self-supervised
  tasks,'' \emph{arXiv preprint arXiv:1904.03416}, 2019.

\bibitem{liu2020mockingjay}
A.~T. Liu, S.-w. Yang, P.-H. Chi, P.-c. Hsu, and H.-y. Lee, ``Mockingjay:
  Unsupervised speech representation learning with deep bidirectional
  transformer encoders,'' in \emph{ICASSP 2020-2020 IEEE International
  Conference on Acoustics, Speech and Signal Processing (ICASSP)}.\hskip 1em
  plus 0.5em minus 0.4em\relax IEEE, 2020, pp. 6419--6423.

\bibitem{schneider2019wav2vec}
S.~Schneider, A.~Baevski, R.~Collobert, and M.~Auli, ``wav2vec: Unsupervised
  pre-training for speech recognition,'' \emph{arXiv preprint
  arXiv:1904.05862}, 2019.

\bibitem{chung2019unsupervised}
Y.-A. Chung, W.-N. Hsu, H.~Tang, and J.~Glass, ``An unsupervised autoregressive
  model for speech representation learning,'' \emph{arXiv preprint
  arXiv:1904.03240}, 2019.

\bibitem{ravanelli2018speaker}
M.~Ravanelli and Y.~Bengio, ``Speaker recognition from raw waveform with
  sincnet,'' in \emph{2018 IEEE Spoken Language Technology Workshop
  (SLT)}.\hskip 1em plus 0.5em minus 0.4em\relax IEEE, 2018, pp. 1021--1028.

\bibitem{devlin2018bert}
J.~Devlin, M.-W. Chang, K.~Lee, and K.~Toutanova, ``Bert: Pre-training of deep
  bidirectional transformers for language understanding,'' \emph{arXiv preprint
  arXiv:1810.04805}, 2018.

\bibitem{kingma2013auto}
D.~P. Kingma and M.~Welling, ``Auto-encoding variational bayes,'' \emph{arXiv
  preprint arXiv:1312.6114}, 2013.

\bibitem{hsu2017unsupervised}
W.-N. Hsu, Y.~Zhang, and J.~Glass, ``Unsupervised learning of disentangled and
  interpretable representations from sequential data,'' in \emph{Advances in
  neural information processing systems}, 2017, pp. 1878--1889.

\bibitem{khurana2019factorial}
S.~Khurana, S.~R. Joty, A.~Ali, and J.~Glass, ``A factorial deep markov model
  for unsupervised disentangled representation learning from speech,'' in
  \emph{ICASSP 2019-2019 IEEE International Conference on Acoustics, Speech and
  Signal Processing (ICASSP)}.\hskip 1em plus 0.5em minus 0.4em\relax IEEE,
  2019, pp. 6540--6544.

\bibitem{li2018disentangled}
Y.~Li and S.~Mandt, ``Disentangled sequential autoencoder,'' \emph{arXiv
  preprint arXiv:1803.02991}, 2018.

\bibitem{uria2016neural}
B.~Uria, M.-A. C{\^o}t{\'e}, K.~Gregor, I.~Murray, and H.~Larochelle, ``Neural
  autoregressive distribution estimation,'' \emph{The Journal of Machine
  Learning Research}, vol.~17, no.~1, pp. 7184--7220, 2016.

\bibitem{oord2016wavenet}
A.~v.~d. Oord, S.~Dieleman, H.~Zen, K.~Simonyan, O.~Vinyals, A.~Graves,
  N.~Kalchbrenner, A.~Senior, and K.~Kavukcuoglu, ``Wavenet: A generative model
  for raw audio,'' \emph{arXiv preprint arXiv:1609.03499}, 2016.

\bibitem{mehri2016samplernn}
S.~Mehri, K.~Kumar, I.~Gulrajani, R.~Kumar, S.~Jain, J.~Sotelo, A.~Courville,
  and Y.~Bengio, ``Samplernn: An unconditional end-to-end neural audio
  generation model,'' \emph{arXiv preprint arXiv:1612.07837}, 2016.

\bibitem{vasquez2019melnet}
S.~Vasquez and M.~Lewis, ``Melnet: A generative model for audio in the
  frequency domain,'' \emph{arXiv preprint arXiv:1906.01083}, 2019.

\bibitem{prenger2019waveglow}
R.~Prenger, R.~Valle, and B.~Catanzaro, ``Waveglow: A flow-based generative
  network for speech synthesis,'' in \emph{ICASSP 2019-2019 IEEE International
  Conference on Acoustics, Speech and Signal Processing (ICASSP)}.\hskip 1em
  plus 0.5em minus 0.4em\relax IEEE, 2019, pp. 3617--3621.

\bibitem{donahue2018adversarial}
C.~Donahue, J.~McAuley, and M.~Puckette, ``Adversarial audio synthesis,''
  \emph{arXiv preprint arXiv:1802.04208}, 2018.

\bibitem{di-gangi-etal-2019-must}
M.~A. Di~Gangi, R.~Cattoni, L.~Bentivogli, M.~Negri, and M.~Turchi,
  ``{M}u{ST}-{C}: a {M}ultilingual {S}peech {T}ranslation {C}orpus,'' in
  \emph{Proceedings of the 2019 Conference of the North {A}merican Chapter of
  the Association for Computational Linguistics: Human Language Technologies,
  Volume 1 (Long and Short Papers)}, 2019, pp. 2012--2017.

\bibitem{ioffe2015batch}
S.~Ioffe and C.~Szegedy, ``Batch normalization: Accelerating deep network
  training by reducing internal covariate shift,'' in \emph{International
  Conference on Machine Learning}, 2015, pp. 448--456.

\bibitem{radford2019language}
A.~Radford, J.~Wu, R.~Child, D.~Luan, D.~Amodei, and I.~Sutskever, ``Language
  models are unsupervised multitask learners.''

\bibitem{bredin2017tristounet}
H.~Bredin, ``Tristounet: triplet loss for speaker turn embedding,'' in
  \emph{2017 IEEE international conference on acoustics, speech and signal
  processing (ICASSP)}.\hskip 1em plus 0.5em minus 0.4em\relax IEEE, 2017, pp.
  5430--5434.

\bibitem{weinberger2009distance}
K.~Q. Weinberger and L.~K. Saul, ``Distance metric learning for large margin
  nearest neighbor classification,'' \emph{Journal of Machine Learning
  Research}, vol.~10, no. Feb, pp. 207--244, 2009.

\bibitem{versteegh2016zero}
M.~Versteegh, X.~Anguera, A.~Jansen, and E.~Dupoux, ``The zero resource speech
  challenge 2015: Proposed approaches and results,'' \emph{Procedia Computer
  Science}, vol.~81, pp. 67--72, 2016.

\bibitem{dunbar2017zero}
E.~Dunbar, X.~N. Cao, J.~Benjumea, J.~Karadayi, M.~Bernard, L.~Besacier,
  X.~Anguera, and E.~Dupoux, ``The zero resource speech challenge 2017,'' in
  \emph{2017 IEEE Automatic Speech Recognition and Understanding Workshop
  (ASRU)}.\hskip 1em plus 0.5em minus 0.4em\relax IEEE, 2017, pp. 323--330.

\bibitem{liu2019mockingjay}
A.~T. Liu, S.~wen Yang, P.-H. Chi, P.~chun Hsu, and H.~yi~Lee, ``Mockingjay:
  Unsupervised speech representation learning with deep bidirectional
  transformer encoders,'' 2019.

\bibitem{lee2012nonparametric}
C.-y. Lee and J.~Glass, ``A nonparametric bayesian approach to acoustic model
  discovery,'' in \emph{Proceedings of the 50th Annual Meeting of the
  Association for Computational Linguistics: Long Papers-Volume 1}.\hskip 1em
  plus 0.5em minus 0.4em\relax Association for Computational Linguistics, 2012,
  pp. 40--49.

\bibitem{ondel2016variational}
L.~Ondel, L.~Burget, and J.~{\v{C}}ernock{\`y}, ``Variational inference for
  acoustic unit discovery,'' \emph{Procedia Computer Science}, vol.~81, pp.
  80--86, 2016.

\bibitem{ebbers2017hidden}
J.~Ebbers, ``Hidden markov model variational autoencoder for acoustic unit
  discovery.'' 2017.

\end{thebibliography}


\end{document}